\documentclass[aps,prb,twocolumn,showpacs,preprintnumbers,amsmath,amssymb,superscriptaddress]{revtex4-1}

\usepackage{graphicx}
\usepackage{color}
\usepackage{dcolumn}
\usepackage{bm}
\usepackage{feynmp}
\usepackage{braket}

\newcommand{\nc}{\newcommand}
\nc{\be}{\begin{eqnarray}}
\nc{\ee}{\end{eqnarray}}
\nc{\bea}{\begin{eqnarray}}
\nc{\eea}{\end{eqnarray}}
\nc{\bean}{\begin{eqnarray*}}
\nc{\eean}{\end{eqnarray*}}
\nc{\mb}{\mbox}
\nc{\rnc}{\renewcommand} 
\nc{\vk}{{\bm k}}
\nc{\vx}{\mb{\bf x}}
\nc{\br}{\mb{\bf r}}
\nc{\bv}{\mb{\bf v}}
\nc{\bp}{\mb{\bf p}}
\nc{\ve}{\mb{\bf e}}
\nc{\vz}{\hat {\mb{\bf z}}}
\nc{\vp}{\mb{\boldmath$p$}}
\nc{\vb}{\mb{\boldmath$b$}}
\nc{\rr}{\mb{\boldmath$r$}}
\nc{\vR}{\mb{\boldmath$R$}}
\nc{\vj}{\mb{\boldmath$j$}}
\nc{\vg}{\mb{\boldmath$g$}}
\nc{\vm}{\mb{\boldmath$m$}}
\nc{\vd}{\mb{\boldmath$d$}}
\nc{\hd}{\mb{\boldmath$\hat{d}$}}
\nc{\vD}{\mb{\boldmath$D$}}
\nc{\vF}{\mb{\boldmath$F$}}
\nc{\vG}{\mb{\boldmath$G$}}
\nc{\vI}{\mb{\boldmath$I$}}
\nc{\vW}{\mb{\boldmath$W$}}
\nc{\x}{\mb{\boldmath$x$}}
\nc{\A}{\mb{\boldmath$A$}}
\nc{\va}{\mb{\boldmath$a$}}
\nc{\vv}{\mb{\boldmath$v$}}
\nc{\vq}{\mb{\boldmath$q$}}
\nc{\vn}{\mb{\boldmath$n$}}
\nc{\vJ}{\mb{\boldmath$J$}}
\nc{\vS}{\mb{\boldmath$S$}}
\nc{\vs}{\mb{\boldmath$\sigma$}}
\nc{\vE}{\mb{\boldmath$E$}}
\nc{\vB}{\mb{\boldmath$B$}}
\nc{\vM}{\mb{\boldmath$M$}}
\nc{\vL}{\mb{\boldmath$L$}}
\nc{\vpsi}{\mb{\boldmath$\psi$}}
\nc{\vphi}{\mb{\boldmath$\varphi$}}
\nc{\Vphi}{\mb{\boldmath$\phi$}}
\nc{\Vomega}{\mb{\boldmath$\Omega$}}
\nc{\ipsi}{\it{\Psi}}
\nc{\vepsilon}{\mb{\boldmath$\epsilon$}}
\nc{\valpha}{\mb{\boldmath$\alpha$}}
\nc{\vgamma}{\mb{\boldmath$\gamma$}}
\nc{\vomega}{\mb{\boldmath$\omega$}}
\nc{\vmu}{\mb{\boldmath$\mu$}}
\nc{\vt}{\mb{\boldmath$\tau$}}
\nc{\vT}{\mb{\boldmath$T$}}
\nc{\vpi}{\mb{\boldmath$\pi$}}
\nc{\nab}{\bm{\nabla}}
\nc{\ov}{\overline}
\nc{\cdott}{\!\cdot\!}
\nc{\cdottt}{\!\!\cdot\!}
\nc{\LL}{\Big{\langle}}
\nc{\RR}{\Big{\rangle}}
\nc{\LR}{\Bigm{|}}
\nc{\vP}{\mb{\boldmath$P$}}
\nc{\nnn}{\nonumber\\}
\nc{\ltsim}{\protect\raisebox{-0.5ex}{$\:\stackrel{\textstyle <}{\sim}\:$}}
\nc{\gtsim}{\protect\raisebox{-0.5ex}{$\:\stackrel{\textstyle >}{\sim}\:$}} 
\nc{\ltsimscript}{\protect\raisebox{-0.5ex}{$\stackrel{\scriptstyle <}{\sim}$}} 
\nc{\gtsimscript}{\protect\raisebox{-0.5ex}{$\stackrel{\scriptstyle >}{\sim}$}} 

\rnc{\figurename}{FIG.}

\nc{\psibar}{\overline{\psi}}
\nc{\cbar}{\overline{c}}
\nc{\intx}{\int d^4x}
\nc{\inty}{\int d^4y}
\nc{\Mhat}{\hat{\bm M}}
\nc{\intk}{\int d^3k}
\nc{\xc}{\bm x_C}
\nc{\kc}{\bm k_C}
\nc{\uk}{u_k^\lambda}
\nc{\rdot}{\dot{r}}
\nc{\kdot}{\dot{k}}
\nc{\ene}{\epsilon_{\bm k}}
\nc{\vf}{v_F}
\nc{\M}{\bm M}

\begin{document}
\title{
Theory of current-driven dynamics of spin textures on a surface of topological insulators
}

\author{Daichi Kurebayashi}
\affiliation{
RIKEN Center for Emergent Matter Science (CEMS), Wako 351-0198, Japan
}
\author{Naoto Nagaosa}
\affiliation{
RIKEN Center for Emergent Matter Science (CEMS), Wako 351-0198, Japan
}
\affiliation{
Department of Applied Physics, University of Tokyo, 7-3-1 Hongo, Bunkyo-ku, Tokyo 113-8656
}

\date{\today}

\begin{abstract}
Spin-transfer torque is one of the important physical quantities to understand for successful application of topological insulators to spintronics.
In this paper, we present analytical expressions of the spin-transfer torques on a surface of a magnetic topological insulator by including the higher-order contributions of momentum, $k^2$-term and the hexagonal warping.
We obtain six different types of the spin-transfer torque including both the field-like and the damping-like torques; the four of them appear only when the higher-order momentum contributions are included.
In addition, we discuss the dynamics of magnetic skyrmions driven by the spin-transfer torques on the surface of the topological insulator.
Unlike the skyrmion dynamics in conventional metals, we find that the dynamics significantly depends on the internal structure of magnetic textures.

\end{abstract}

\maketitle

\section{Introduction}
Electrical manipulation of magnetic textures is of major interest in the field of spintronics for applications to low-energy consumption electronic devices\cite{Zutic2004}.
In conventional ferromagnetic metals, a spin-transfer torque is widely used to control magnetic textures\cite{Berger1996,Slonczewski1996}.
The racetrack memory utilizing magnetic domain walls and the spin-transfer torque, for instance, is proposed as a new spintronics devices replacing conventional electronic memories\cite{Parkin2008a}.
The requirement of a considerable current to operate magnetic domain walls due to the pinning, however, impedes a commercial application of the racetrack memory.
To overcome this technical challenge, a magnetic skyrmion has attracted much interests.
The magnetic skyrmion is a two-dimensional swirling spin texture, recently discovered in the systems with strong spin-orbit coupling such as chiral-lattice magnets with the Dzyalosinskii-Moriya interaction\cite{Bogdanov2001,Page2009,Munzer2010,Kanazawa2010,Schulz2012,Nagaosa2013}.
Because they can move around defects and avoid pinning, the skyrmions have a much lower threshold current to drive than that of the domain walls.
The skyrmions have other advantages in application such as the thermal stability due to the topological property and the capability to make the high integration density devices\cite{Fert2013,Kang2016}.

As an alternative route to achieve the high-performance devices, utilizing the topological properties of materials have been one of main interests known as the topological spintronics\cite{Pesin2012,Smejkal2018}.
A topological insulator is a promising candidate, whose surface states have strong correlation between spin and currents known as spin-momentum locking\cite{Hasan2010,Qi2011}.
Due to the spin-momentum locking, it is expected that more efficient magnetic manipulation can be achieved in the magnetic topological insulator, and electric manipulation of the magnetic texture\cite{Nomura2010,Hurst2015,Andrikopoulos2016}, spin-charge conversion \cite{Shiomi2014,Kondou2016a}, and magnetization switching by spin-orbit torque\cite{Mellnik2014,Fan2014,Yasuda2017} have been proposed and experimentally performed.

Recently, in a heterostructure consisting of a topological insulator and a magnetically-doped topological insulator, the evidence of skyrmion formation has been observed by the Hall measurement\cite{Yasuda2016}.
They observe a deviation from the conventional anomalous Hall effect signal, attributing it to the geometric Hall effect arising from the emergent electromagnetic field of the skyrmion spin texture.
The skyrmion on the surface of the topological insulator is interesting from the viewpoints of not only the interplay of the topology both in real and momentum spaces, but also its potential to provide more efficient means to control magnetic textures.


In order to understand dynamics of the magnetic texture, the spin torque is an important physical quantity.
There have been theoretically shown that the spin torque on the surface of the topological insulator is modified greatly from that of conventional ferromagnetic metals\cite{Garate2010,Yokoyama2010,Sakai2014,Ndiaye2015}.
The one peculiar feature of the surface of the topological insulator is that the absence of conventional spin-transfer torque due to the one-to-one correspondence between current and spin operators, $\hat{\bm j} \propto \hat{z}\times \hat{\bm \sigma}$\cite{Sakai2014}.
The correspondence is a consequence of considering only the $k$-linear Hamiltonian, or the spin-momentum locking.
However, it have been shown by angle-resolved photo emission spectroscopy experiments that the band dispersion deviates from the Dirac cone as increasing the Fermi energy in a typical topological insulator such as $\rm Bi_2Se_3$\cite{Hsieh2009a,Kuroda2010}.
The deviation can be explained by higher-order contributions of momentum, i.e. a $k^2$-term breaking the particle-hole symmetry and a hexagonal warping term modifying the Fermi surface to a hexagon\cite{Fu2009}.
Those higher-order contributions of momentum break the one-to-one correspondence between current and spin operators, therefore expected to affect the spin-transfer torque significantly.
However, the studies on the effect of the $k^2$-term and the warping term to the spin-transfer torque have not been conducted yet, and highly demanded to understand the dynamics of magnetic textures on the surface of the topological insulator. 

In this paper, we analytically derive spin-transfer torques on the surface of the topological insulator including $k^2$-term and the warping term.
Then, we discuss the dynamics of skyrmion driven by the obtained spin-transfer torques as an expample.

\section{Spin torques}
\begin{figure}[tbp]
\centering
\includegraphics[width = 0.9\linewidth]{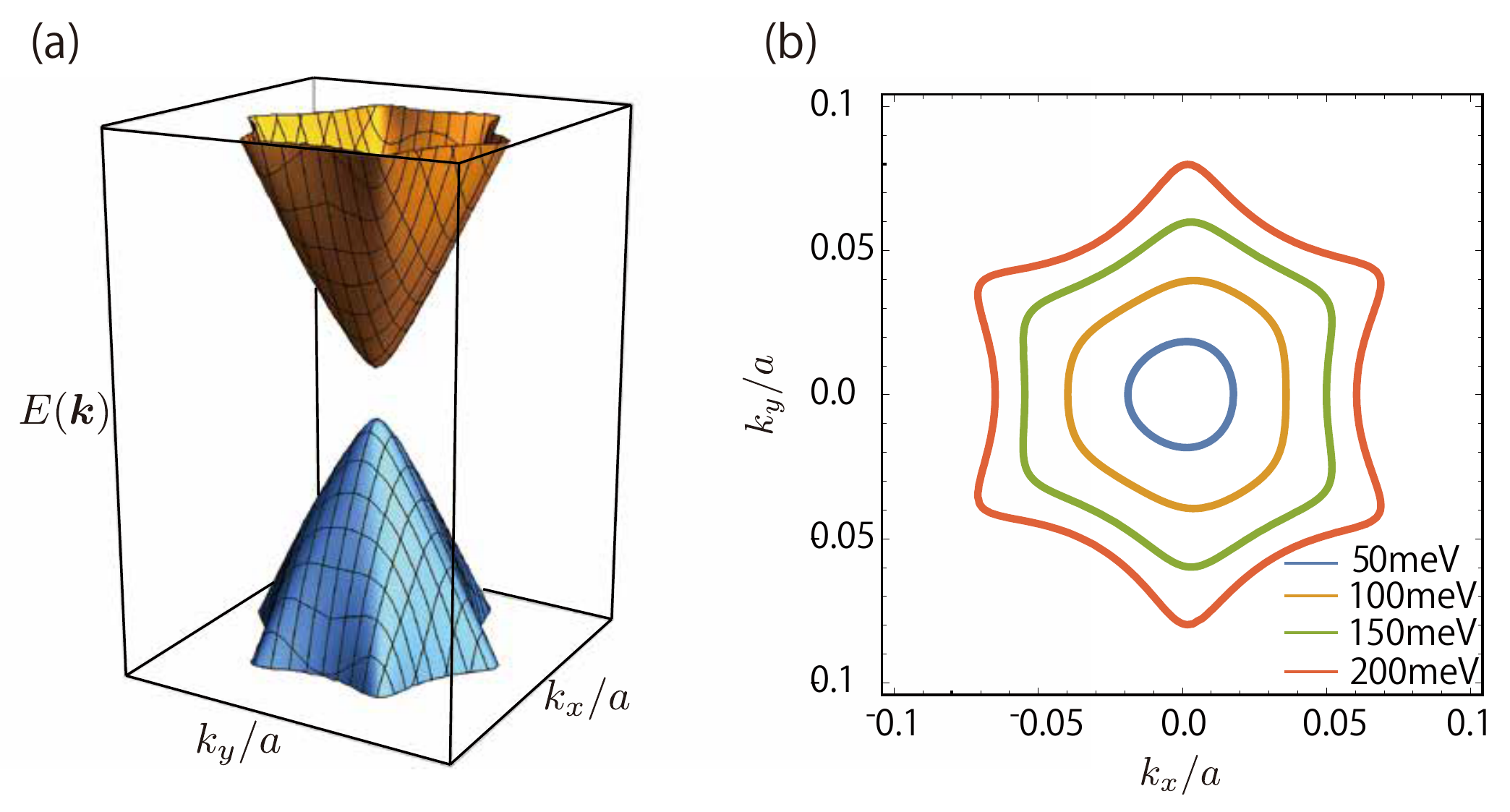}
\caption{
(color online) Band structure and energy contour of the surface state of magnetic topological insulator.
}
\label{band}
\end{figure}

In this study, we consider the two band Weyl Hamiltonian with $k^2$-term, the warping term and exchange coupling to local magnetic moments.
The low energy effective Hamiltonian\cite{Fu2009} is given by $H_D = \int d^2k \psi^\dagger_{\bm k}\mathcal{H}_{D}(\bm k)  \psi_{\bm k}$ where
\bea
\nonumber \mathcal{H}_{D}(\bm k) &=& v_F \left(k_x \sigma_y-k_y \sigma_x\right) + JM_z\sigma_z\\
&&+\gamma k^2 \sigma_0 +  \frac{\lambda}{2} \left(k_+^3+k_-^3\right)\sigma_z ,
\label{H0}
\eea
$\psi_{\bm k} = \left(\psi_{\bm k\uparrow},\psi_{\bm k\downarrow}\right)^T$ is an electron annihilation operator with the wave vector $\bm k$, $v_F$ is Fermi velocity, $\sigma_i$ is the Pauli matrix representing the real spin degrees of freedom, $J$ is the exchange coupling between itinerant electron's spin and local magnetic moments ($J<0$), $M_z$ is the amplitude of $z$-component of magnetization, and $k_\pm = k_x\pm i k_y$.
The third term describes a quadratic term breaking the particle-hale symmetry, while the fourth term describes the hexagonal warping term.
The quadratic and the warping terms are characterized by the coefficient $\gamma$ and $\lambda$, respectively.
The eigenvalues of the Hamiltonian is given by
\bea
\varepsilon^s_{\bm k} &=& \gamma k^2 + s\sqrt{v_F^2k^2 + \left(JM_z + \lambda k^3\cos3\theta_{\bm k} \right)^2}
\eea
where $s = \pm1$ labels the conduction and the valence bands, $\theta_{\bm k} = \tan^{-1}(k_y/k_x)$, and the dispersion is depicted in FIG.\ref{band}.
As the perturbation, we introduce the exchange coupling to transverse fluctuations of magnetization and the electromagnetic coupling given as
\bea
H'&=& J \int d^2r \ \bm u(\bm r) \cdot \psi^\dagger_{\bm r}\bm \sigma \psi_{\bm r}\\
H''&=& -e\int d^2k e^{i\omega t} \bm A(\bm k,\omega)\cdot  \psi^\dagger_{\bm k} \bm v_{\bm k} \psi_{\bm k}
\eea
where $\psi_{\bm r} = (\psi_{\bm r\uparrow}, \psi_{\bm r\downarrow})^T$ is an electron annihilation operator in the Wannier state centered at $\bm r$, $\bm u(r)$ is in-plane component of the magnetization, and a group velocity is given by $\bm v_{\bm k} = \frac{\partial H_D}{\partial \hbar \bm k}$.
As for impurities, we consider non-magnetic impurities with short-range potential given as
$
\hat{V}_{\rm imp} = \int d\bm r\psi^\dagger_{\bm r}V_{\rm imp}(\bm r)\psi_{\bm r}
$ 
where $V_{\rm imp}(\bm r) = u_0\sum_i \delta(\bm r - \bm R_i)$.
A Gaussian average is taken for impurity positions as $\braket{V_{\rm imp}(\bm r)V_{\rm imp}(\bm r')} = n_i u_0\delta(\bm r-\bm r')$ where $n_i$ is the concentration of impurities.

The Matsubara Green's function of the Hamiltonian Eq.(\ref{H0}) is given as
\bea
\hat{G}(i\omega_n,\bm k) = \left[i\omega_n + E_F - \mathcal{H}_D(\bm k) - \Sigma(i\omega_n,\bm k)\right]^{-1}
\eea
where $E_F$ is the Fermi energy and $\Sigma(i\omega_n,\bm k)$ is the self-energy induced by impurity scattering.
In the first Born approximation, the imaginary part of the self-energy is geven as ${\rm Im\ } \Sigma = n_i u_0^2\sum_{\bm q} G(i\omega_n,\bm q) = \pi n_i u_0^2 D(E_F) {\rm sgn}(\omega_n) \equiv \frac{\hbar}{2\tau} {\rm sgn}(\omega_n)$ where $D(E_F)$ is the density of states at the Fermi energy and $\tau$ is the scattering lifetime.
We neglect a correction to the mass term, the part of the self-energy coupled to $\sigma_z$.
In the following, we assume $1 << E_F\tau/\hbar$ corresponding to a weak impurity scattering regime and calculate spin torque in the lowest order of $1/\tau$.

\begin{figure*}[tbp]
\begin{center}
\includegraphics[width = 0.8\linewidth]{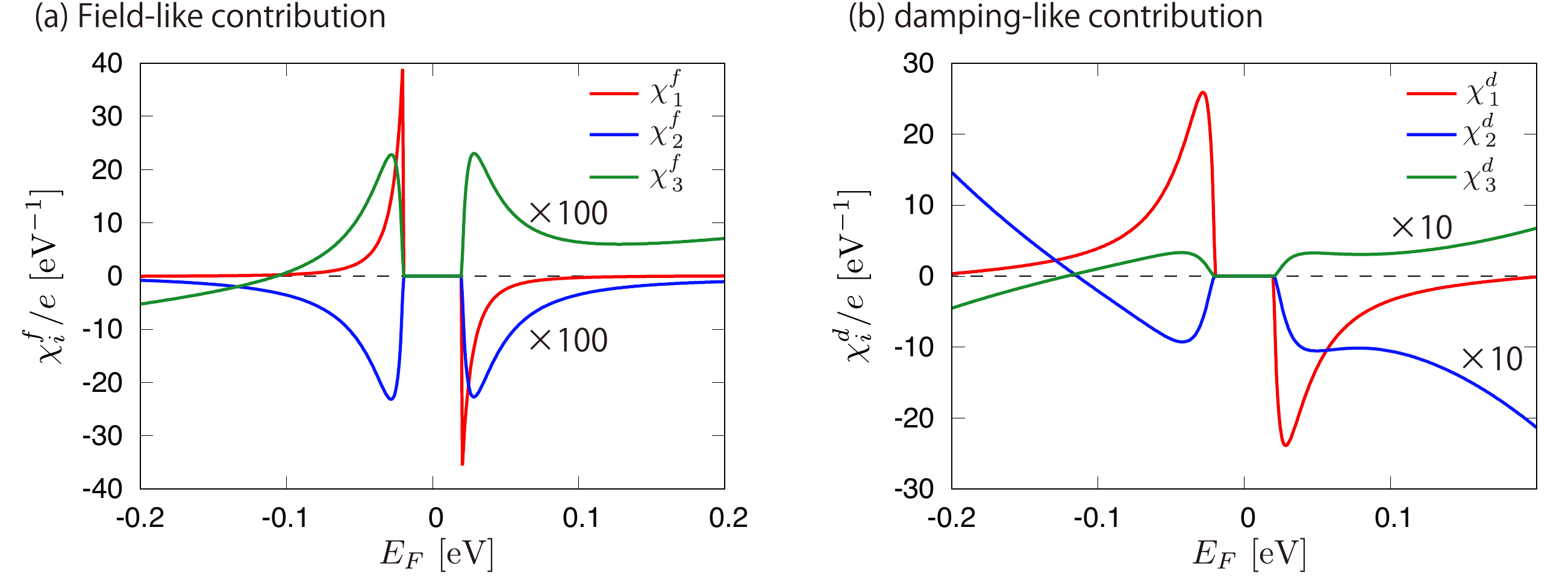}
\caption{
(color online) Coefficients corresponds to (a) the field-like and (b) the damping-like spin-transfer torques.
Parameters are taken as $v_F = 2.5$ eV$\AA$, $\gamma = 7.2$ eV$\AA^2$, $\lambda = 250$ eV$\AA^3$, and $JM_z = 0.02$ eV.
}
\label{fig2}
\end{center}
\end{figure*}

Magnetization dynamics is described by the Landau-Lifshitz-Gilbert (LLG) equation,
\bea
\frac{d\M}{dt} = \gamma_0 \bm B\times \M + \frac{\alpha}{S} \M\times\frac{d\M}{dt} + T_e
\eea
where $\gamma_0$ is the gyromagnetic ratio, $\bm B$ is an effective magnetic field, $S$ is the amplitude of the local magnetic moments, and $\alpha$ is the damping constant.
The last term is known as spin torques emerging from coupling between local moments and itinerant electrons, which is defined as
\bea
\bm T_e = \frac{J}{\hbar \rho_s} \M \times \braket{\bm \sigma}
\label{torque}
\eea
where $\rho_s$ is the number of local magnetic elements per unit volume, and $\langle \bm \sigma \rangle$ is a non-equilibrium spin density of the itinerant electrons.
Namely, spin torques can be evaluated by calculating electrically-induced spin density.

The spin density induced by an electric field within linear response theory as
\bea
\braket{\sigma_i(\bm q)} &=& \chi_{ij}(\bm q) E_j\\
\chi_{ij}(\bm q) &=& \lim_{\Omega\rightarrow 0} \frac{K_{ij}(\Omega+i0,\bm q)-K_{ij}(0,\bm q)}{i\Omega}
\label{chi}
\eea
where $K_{ij}(\Omega,\bm q) = i \int_0^\infty dt e^{i\Omega t}\braket{[\sigma_i(t, \bm q), j_j]}$ is a dynamical correlation function between the electron spin and the electric currents.
Here we consider a uniform electric field, thus the wave vector $\bm q$ comes only from the magnetization fluctuation.
We further assume that the spatial variation of the magnetization fluctuation is small and expand the response function up to linear order in $\bm q$ and $\bm u$ as
\bea
K_{ij}(\Omega,\bm q) \approx K_{ij}(\Omega) + K_{ijkl}(\Omega) q_l u_k
\eea
where the uniform contribution corresponds to the spin-orbit torque while the spatial dependent contribution corresponds to the spin-transfer torque.

The spin-orbit torque on the surface of the topological insulator has been well studied\cite{Yokoyama2010,Sakai2014}, and the spin density corresponds to the spin-orbit torque is given by
\bea
\braket{\bm \sigma} = \chi_{\rm sot}^f \hat{\bm z}\times \bm E + \chi_{\rm sot}^dM_z \bm E
\label{sot}
\eea
where the field-like contribution is attributed to the Rashba-Edelstein effects due to the spin-momentum locking on the surface of the topological insulator\cite{Wang2011a} and the damping-like contribution arises from the electromagnetic coupling\cite{Garate2010,Ndiaye2017}.
It is worthy to note that the damping-like contribution is proportional to the anomalous Hall conductivity, reflecting the topological property of two-dimensional massive Dirac fermions.
Furthermore, this contribution becomes finite even when the fermi energy is located in the exchange gap, $|E_F|<|JM_z|$.


\begin{figure*}[tbp]
\centering
\includegraphics[width=0.8\linewidth]{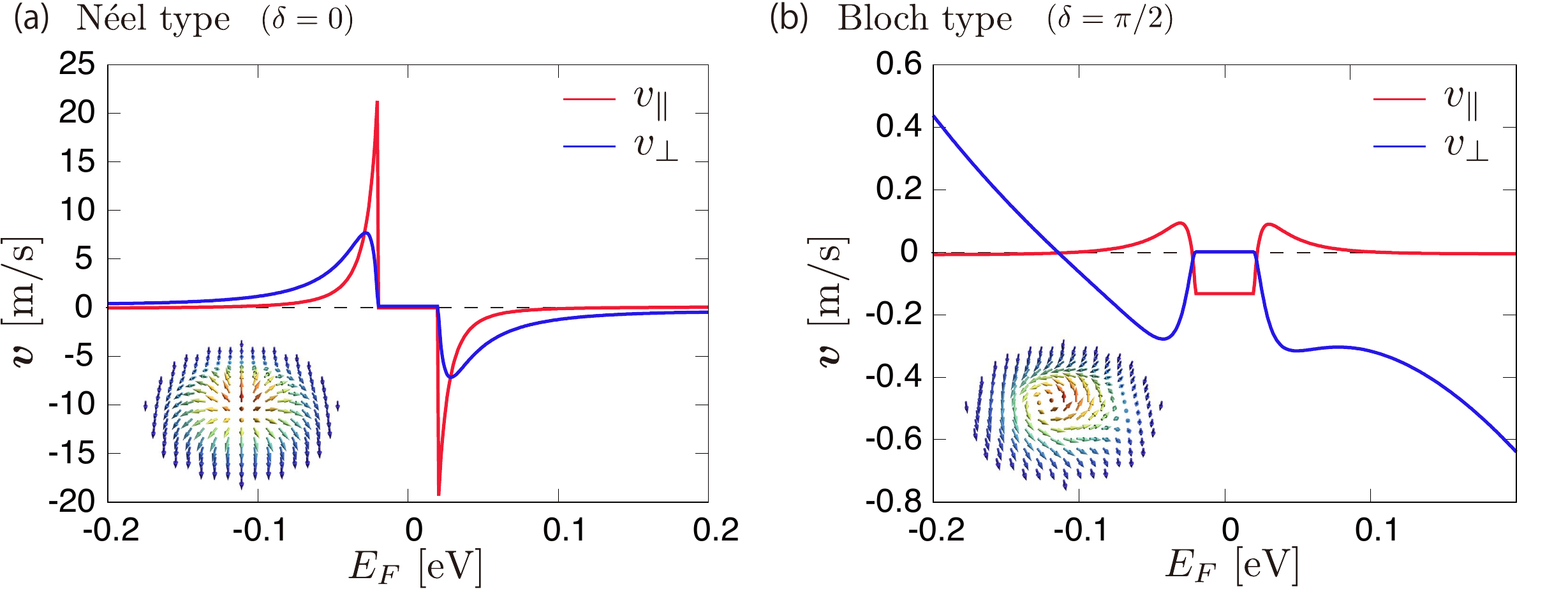}
\caption{
(color online) Drift velocity of (a) the N\'eel type skyrmion and (b) the Bloch type skyrmion.
Parameters are taken as $v_F = 2.5$ eV$\AA$, $\gamma = 7.2$ eV$\AA^2$, $\lambda = 250$ eV$\AA^3$, $m_0\equiv JM_z = 0.02$ eV, the size of the skyrmion $\sim 20$nm, and the applied electric field $|\bm E| = 10^4$ m/V.
}
\label{fig_skyrm}
\end{figure*}

The correlation function corresponding to the spin-transfer torque, which is our main focus in this paper, is evaluated by 
\bea
\nonumber K_{ijkl}(\Omega_m) &=&\frac{-eJ}{\beta} \sum_{n,\bm k} {\rm Tr}\left[ \sigma_i \hat{G}^-v_j\hat{G}\sigma_k \hat{G}v_l\hat{G}\right. \\
&&\hspace{1cm }\left. -\sigma_i \hat{G}v_l\hat{G}\sigma_k \hat{G}v_j\hat{G}^+\right]
\eea
where $\hat{G}$ and $\hat{G}^\pm$ denote $\hat{G}(i\omega_n, \bm k)$ and $\hat{G}(i\omega_n\pm i\Omega_m,\bm k)$, respectively.
The leading order of the spin response function associated with the STT in damping rate $1/\tau$ is calculated in the Appendix and, as a result, the spin density induced by the applied field is given by
\bea
\nonumber \braket{\bm \sigma} &=& \chi_{1}^f M_z \hat{\bm z} \times \left(\bm \nabla \cdot \M\right)\bm E + \chi_{1}^d \left(\bm \nabla \cdot \M \right)\bm E\\
\nonumber&&+ \chi_{2}^f M_z \hat{\bm z} \times \left(\bm E \cdot \bm \nabla\right)\M + \chi_{2}^d \left(\bm E\cdot \bm \nabla \right)\M\\
&& +\chi_{3}^f M_z \hat{\bm z} \times \bm \nabla \left(\bm E \cdot \M \right) +\chi_{3}^d\bm \nabla \left(\bm E \cdot \M \right)
\label{stt}
\eea
where terms proportional to $\chi_{1}^f$, $\chi_{2}^f$, and  $\chi_{3}^f$ correspond to the field-like contributions and rests correspond to the damping-like contributions.
When $|E_F|>>|JM_z|$, the coefficients are obtained as
\bea
\chi_{i}^f &=& -\frac{eJ^2\tau}{2\pi} {\rm sgn}(E_F) \Lambda_i^f(E_F)\\
\chi_{i}^d &=& \frac{eJ\tau^2}{4\pi} {\rm sgn}(E_F) \Lambda_i^d(E_F)
\eea
where ${\rm sgn(x)}$ is the sign function, and the functions characterizing the field-like contribution are$\Lambda_1^f = \frac{1}{E_F^2}- \lambda^2\frac{31E_F^4}{8 v_F^4}$, $\Lambda_2^f = \frac{\gamma}{E_Fv_F^2}+ \lambda^2\frac{E_F^2}{8v_F^6}$, $\Lambda_3^f = -\frac{\gamma }{E_Fv_F^2} - \lambda^2\frac{31E_F^2}{8 v_F^6}$, and those of the damping-like contribution are $\Lambda_1^d = 1 -\gamma\frac{E_F}{2v_F^2}-\lambda^2\frac{2E_F^4}{ v_F^6}$, $\Lambda_2^d = \gamma\frac{3E_F}{2v_F^2}+\lambda^2\frac{9E_F^4}{2v_F^6}$, $\Lambda_3^d = -\gamma\frac{E_F}{2v_F^2}	- \lambda^2\frac{3E_F^4}{2v_F^6}$.
The full expression of the coefficients are presented in the Appendix.
The spin-transfer torque is, then, obtained by Eq.(\ref{torque}) as $\bm T_e = (J/\hbar \rho_S)\bm M \times \braket{\bm \sigma}$.
It is first noted that the first two terms becomes finite in the absence of the $k^2$-term and the warping term, whereas the other four terms appear only when the deviation from the linear dispersion is present.
The first two contributions can be understood by the spin-transport correspondence\cite{Sakai2014}.
In the absence of the $k^2$-term and the warping-term, the current operator and the spin operator are identical except for a constant factor,
$
\bm j = -ev_F\hat{z}\times \bm \sigma.
$
The electron spins couple to the in-plane magnetization by the exchange interaction as $-\bm M\cdot \bm \sigma$, while the current couple to the vector potential as $-e\bm A\cdot \bm j$.
Due to the correspondence between the current operator and the spin operator, the vector potential and in-plane magnetization couple to electrons in the same manner, namely the electrons cannot distinguish the perturbations.
By using this correspondence, the first two terms of Eq.(\ref{stt}) can be interpreted by the current density as
\bea
\braket{\bm j} = -ev_F\left[\chi_{1}^f M_z\hat{z}\times B_z\bm E + \chi_{1}^dB_z\bm E\right],
\eea
where $B_z$ is a $z$-component of the magnetic field.
This current density is nothing but the ordinal Hall current, therefore the spin torque corresponding to the effect is attributed to the Hall effect-driven spin torque.
Actually, the Hall effect-driven spin torque is a only gauge-invariant form of spin torque in the first order of spatial derivative of magnetization.

Contrary, the other four contributions given in Eq.(\ref{stt}) are interpreted in the current transport picture as
\bea
\nonumber \braket{\bm j} &=& -ev_F \left[\chi_{2}^f M_z \hat{\bm z} \times \left(\bm E \cdot \bm \nabla\right)\bm A + \chi_{2}^d \left(\bm E\cdot \bm \nabla \right)\bm A\right.\\
&&\left. \hspace{0.5cm} +\chi_{3}^f M_z \hat{\bm z} \times \bm \nabla \left(\bm E \cdot \bm A \right) +\chi_{3}^d\bm \nabla \left(\bm E \cdot \bm A \right)\right]
\eea
which explicitly depends on the vector potential, therefore must vanish because of the U(1) gauge symmetry.
Due to the spin-transport correspondence, the spin density corresponding to the currents becomes also zero when the $k^2$-term and the warping term are absent.
Once the $k^2$-term and the warping-term are included, on the other hand, the correspondence is no longer present and the other form of the spin-transfer torque is allowed.
It is, indeed, shown that the coefficients become finite only when the $k^2$-term and the warping-term are included.
Note that the spin torques characterized by the coefficients $\chi_{2}^f$ and $\chi_{2}^d$ coincide with the conventional adiabatic and non-adiabatic spin-transfer torques, respectively\cite{Slonczewski1996,Berger1996}.
The coefficients are shown in FIG.\ref{fig2} with typical parameters for $\rm Bi_2Se_3$\cite{Liu2010,Kim2016b}.
Although the Hall effect-driven spin torques are dominant in both the field-like and the damping-like spin torques, the other terms including the conventional spin-transfer torques give certain contributions.
The chemical potential dependence also differs between the Hall effect-driven spin torque and the conventional spin-transfer torque.
The Hall effect-driven spin torques change its sign depending on the carrier types, whereas the sign of the conventional spin-transfer torques are independent of the carrier type in the vicinity of the exchange gap.

\section{Skyrmion dynamics}
We, then, discuss the skyrmion dynamics driven on the surface of the topological insulator.
The steady motion of the stabilized skyrmion is described by the Thiele equation\cite{Thiele1973} as
\bea
\left(G_{ij} - \alpha D_{ij}\right)v_j = F_i
\label{thiele}
\eea
where $\bm v$ is the drift velocity of the skyrmion,
$G_{ij}$ is the Magnus force tensor, $D_{ij}$ is the dissipative force tensor, and 
\bea
F_i = \frac{1}{4\pi} \int d^2 r \frac{J}{\rho_s \hbar}\partial_i \bm M\cdot \braket{\bm \sigma}
\eea
is the external force exerted by the spin torque.
In addition to the spin accumulation corresponding to the spin-transfer torque, Eq.(\ref{stt}), we included spin-orbit contributions given Eq.(\ref{sot}).
An azimuthally-symmetric skyrmion is parameterized by\\
$
\bm M(\bm r) = \left[
\cos (\phi+\delta) \sin\theta(\bm r),
\sin (\phi+\delta) \sin\theta(\bm r),
\cos\theta(\bm r)\right]^T
$
where $r$ and $\phi$ are the polar coordinates, $\delta$ is the helicity of the skyrmion, and $\theta(\bm r)$ determine a radial profile.
The skyrmion is called the N\'eel type when $\delta = 0, \pi$ and the Bloch type $\delta = \pm\pi/2$.
With the parameterization, the gyromagnetic coupling and the dissipative tensors are respectively calculated as
$G_{ij} = \mathcal{Q}\epsilon_{ij}$ and $D_{ij} = \mathcal{D}\delta_{ij}$ where $\mathcal{Q}\in \mathbb{Z}$ is the skyrmion number.
Also, the force terms are evaluated as $F_i = \left[q_\parallel \delta_{ij}+q_\perp\epsilon_{ij}\right]E_i$ where
\bea
\nonumber q_\parallel &=& \frac{J}{\hbar \rho_S}\left[\mathcal{C}\chi_{SOT}^d\cos\delta\right.\\
&&\left.+\mathcal{D}\left(\chi_{1}^d\cos^2\delta+\chi_{2}^d+\chi_{3}^d\cos^2\delta\right)\right],\\
\nonumber q_\perp &=& \frac{J}{\hbar\rho_S} \left[
\mathcal{C}\chi_{SOT}^d \sin\delta
-\bar{\mathcal{D}} \left(\chi_{1}^f\cos^2\delta+\chi_{3}^f\sin^2\delta\right)\right.\\
&&\left.-\frac{2}{3}\mathcal{Q} \left(\chi_{1}^f+\chi_{3}^f\right)\cos^2\delta+\mathcal{Q}\chi_{2}^f
\right],
\eea
where $
\mathcal{C} \equiv \frac{1}{4}\int_0^\infty dr r\left[\left(\frac{\partial \theta}{\partial r}\right)\cos\theta + \frac{\sin\theta(\bm r)}{r}\right]
$
and
$
\bar{\mathcal{D}} \equiv \frac{1}{4}\int_0^\infty dr r\cos\theta\left[\left(\frac{\partial \theta}{\partial r}\right)^2 + \frac{\sin^2\theta(\bm r)}{r^2}\right].
$
Note that the field-like spin-orbit torque does not contribute to the skyrmion dynamics.
Finally, the drift velocity is derived from Eq.(\ref{thiele}) as
\bea
\bm v_\parallel &=& \frac{\mathcal{Q}q_\perp - \alpha\mathcal{D}q_\parallel}{\mathcal{Q}^2+(\alpha\mathcal{D})^2}  \bm E\\
\bm v_\perp &=& -\frac{\mathcal{Q}q_\parallel + \alpha\mathcal{D}q_\perp}{\mathcal{Q}^2+(\alpha\mathcal{D})^2}\left(\hat{z} \times \bm E\right).
\eea
With the typical parameters for $\rm Bi_2Se_3$, the drift velocity of the N\'eel and the Bloch skyrmions are presented in FIG.\ref{fig_skyrm}.
It is shown that the N\'eel type skyrmion moves one order of magnitude faster than the Bloch type skyrmion.
This can be understood by the difference in the spin torque whose responsible to the dynamics.
Since the N\'eel type skyrmion has divergence in the magnetic structure, the Hall effect-driven torque gives a dominant contribution to the velocity.
Contrary, because of a divergence-less texture, the Hall effect-driven torque is vanishing in the Bloch type skyrmion.
Instead, the conventional spin-transfer torque dominantly contributes to the dynamics of the Bloch type skyrmion.
The chemical potential dependence of the velocity also differs.
For the N\'eel type skyrmion, the velocity changes its sign depending on carrier type, whereas does not for the Bloch type skyrmion in the low doped regime ($|E_F|<0.1$ eV).

It has been theoretically proposed that the type of skyrmion on the surface of the topological insulator depends on the chemical potential: the Bloch skyrmion is preferred for the low-doping case whereas the N\'eel type skyrmion is stabilized when the chemical potential is away from the exchange gap\cite{Wakatsuki2015}.
There also exists regular magnetic interactions contributing to the chirality of the skyrmion such as the geometric Dzyaloshinskii-Moriya (DM) interaction and the dipole-dipole interaction.
The N\'eel type skyrmion, for instance, is stabilized by the interfacial DM interaction in a thin film geometry, while the Bloch type skyrmion is stabilized by the bulk DM interaction and the dipole-dipole interaction.
The determination of the chirality is important for the device application of skyrmion.
The chirality of the skyrmion on the surface of the topological insulator, however, has yet not observed.
Our result suggests that we can determine the type of the skyrmion by observing velocity while tuning the Fermi energy by the gate control\cite{Yu2017a,Montoya2017}.


As another mechanism to manipulate magnetic textures on the surface of the topological insulator, utilizing a localized charge accumulation at textures has been theoretically proposed\cite{Nomura2010,Hurst2015a,Andrikopoulos2016}.
When the Fermi energy is located in the exchange gap, $|E_F|<|JM_z|$, the magnetic texture on the surface of the topological insulator accompanies an additional charge density due to the chiral edge states.
Because of the additional charge accumulation attached to the textures, it is possible to control the magnetic textures by the external electric field.
This mechanism is consistent with the driving mechanism caused by the damping-like spin-orbit torque, Eq.(\ref{sot}), reflecting the nature of the quantum anomalous Hall insulator.
However, for the doped case where $|E_F|>|JM_z|$, this mechanism is assume to be small due to the screening effect.
In this doped regime, we expect that the current-induced spin-transfer torque, Eq.(\ref{stt}), becomes the leading driving mechanism.


In conclusion, we theoretically investigate the spin-transfer torque on the surface of the topological insulator by including the higher order contributions of momentum, $k^2$-term and the warping term.
By using the linear response theory, we derive the analytical expression of the spin-transfer torque and obtain six different types of the spin-transfer torques; the four of them appear as a consequence of the higher order momentum contributions.
We further consider the dynamics of magnetic textures driven by the obtained spin-transfer torques.
As a result, we predict that the skyrmion dynamics significantly differs depending on the internal structure of the skyrmion.
We hope that our theory gives qualitative understanding of magnetic textures on the surface of the topological insulator and helps achieving next generation spintronics devices with the topological insulator.

The authors are grateful to K. Yasuda for insightful discussions.
D.K. was supported by the RIKEN Special Postdoctoral Researcher Program.
N.N. was supported by JST CREST Grant Number JPMJCR1874 and JPMJCR16F1, Japan, and JSPS KAKENHI Grant numbers 18H03676 and 26103006.

\appendix*
\section{Details of calculation}
\begin{widetext}
In this Appendix, we give a detail calculation for the spin-transfer torques. 
The response function associated with the spin-transfer torque is separated into the field-like and the damping-like contributions as
\bea
\nonumber K_{ijkl}(\Omega_m) &=&\frac{-eJ}{\beta} \sum_{n,\bm k}\left( {\rm Tr}\left[ \sigma_i \hat{G}(\omega_n - \Omega_m,\bm k)v_k\hat{G}(\omega_n ,\bm k)\sigma_j \hat{G}(\omega_n,\bm k )v_l\hat{G}(\omega_n ,\bm k)\right]\right. \\
&&\hspace{1cm }\left. - {\rm Tr}\left[ \sigma_i \hat{G}(\omega_n ,\bm k)v_l\hat{G}(\omega_n ,\bm k)\sigma_j \hat{G}(\omega_n,\bm k )v_k\hat{G}(\omega_n + \Omega_m,\bm k)\right]\right)\\
&\equiv&K_{ijkl}^{d}(\Omega_m) +JM_zK_{ijkl}^{f}(\Omega_m) 
\eea
Let us first evaluate the field-like contributions.
The field-like contribution can be evaluated as
\bea
\nonumber JM_zK_{ijkl}^{f} &\approx& \frac{-ieJ}{\beta} \sum_{n}\int_0^\infty\frac{kdk}{2\pi}\left\{{\rm Im}[A_{+}]\left[g_+(\omega_n-\Omega_m,k)g_+(\omega_n, k)^2g_-(\omega_n, k)+g_+(\omega_n+\Omega_m, k)g_+(\omega_n, k)^2g_-(\omega_n,k)\right]\right.\\
&&+{\rm Im}[B_{+}]\left[g_+(\omega_n-\Omega_m, k)g_+(\omega_n, k)g_-(\omega_n, k)^2+g_+(\omega_n+\Omega_m, k)g_+(\omega_n, k)g_-(\omega_n, k)^2\right]+(+\leftrightarrow-)\\
\nonumber &\approx&-\frac{eJ^2M_z\Omega\tau}{2\pi E_F^2}{\rm sgn}(E_F)\Theta(E_F^2-m_0^2)\left[ \epsilon_{ik}\delta_{ij}\left(1- \gamma \frac{2m_0^2}{E_F v_F^2} - \lambda^2\frac{(E_F^2-m_0^2)^3(31E_F^2+9m_0^2)}{8E_F^4 v_F^6}\right)\right.\\
\nonumber&&\hspace{3cm}+\epsilon_{ij}\delta_{kl}\left(\gamma \frac{E_F^2-m_0^2}{E_Fv_F^2}+ \lambda^2\frac{(E_F^2-m_0^2)^2(E_F^6-34m_0^2E_F^4-35m_0^4E_F^2-40m_0^6)}{8E_F^6 v_F^6}\right)\\
&&\hspace{3cm}\left . -\epsilon_{il}\delta_{jk}\left(\gamma \frac{E_F^2-m_0^2}{E_Fv_F^2}+ \lambda^2\frac{(E_F^2-m_0^2)^2(31E_F^6-10m_0^2E_F^4-17m_0^4E_F^2-40m_0^6)}{8E_F^6 v_F^6}\right)\right]
\eea
where
$
g_s (\omega_n,\bm k) = \left[i\omega_n  + E_F- \gamma k^2 -s \varepsilon_D + i \frac{1}{2\tau}(E_F) {\rm sgn}(\omega_n)\right]^{-1}
$, $m_0 = JM_z$ is the exchange gap, $\Theta(x)$ is the step function, and $A_s$ and $B_s$ are the matrix elements defined as
\bean
A_s &\equiv& \int_0^{2\pi}\frac{d\theta}{2\pi} \left[\braket{s|v_k|s}\braket{s|\sigma_j|s}\braket{s|v_l|\bar{s}}\braket{\bar{s}|\sigma_i|s}+\braket{s|v_k|s}\braket{s|\sigma_j|\bar{s}}\braket{\bar{s}|v_l|s}\braket{s|\sigma_i|s}+\braket{s|v_k|\bar{s}}\braket{\bar{s}|\sigma_j|s}\braket{s|v_l|s}\braket{s|\sigma_i|s}\right],\\
B_s &\equiv& \int_0^{2\pi}\frac{d\theta}{2\pi} \left[\braket{s|v_k|s}\braket{s|\sigma_j|\bar{s}}\braket{\bar{s}|v_l|\bar{s}}\braket{\bar{s}|\sigma_i|s}+\braket{s|v_k|\bar{s}}\braket{\bar{s}|\sigma_j|s}\braket{s|v_l|\bar{s}}\braket{\bar{s}|\sigma_i|s}+\braket{s|v_k|\bar{s}}\braket{\bar{s}|\sigma_j|\bar{s}}\braket{\bar{s}|v_l|s}\braket{s|\sigma_i|s}\right].
\eean
To perform the momentum integral, we simplified calculation by replacing the energy with the angle-averaged energy as
\bea
\bar{\varepsilon}_D(k) \equiv \int_0^{2\pi} \frac{d\theta}{2\pi}\sqrt{v_F^2k^2+(m_0+\lambda k^3 \cos\theta)^2}\approx\sqrt{v_F^2k^2+m_0^2} + \lambda^2 \frac{v_F^2 k^8 }{(v_F^2k^2+m_0^2)^{3/2}} = \varepsilon_0(k) +\lambda^2\frac{v_F^2k^8}{\varepsilon_0(k)^3}.
\eea
We also assume the weak scattering regime, $1<<E_F\tau$, thus we have the relationship
$
g_s^R - g_s^A \approx -2\pi i \delta(E_F - \gamma k^2 - s\bar{\varepsilon}_D).
$

Similarly, the damping-like contribution can also be evaluated as
\bea
\nonumber K_{ijkl}^d &\approx& \frac{-eJ}{\beta}\sum_s\int_0^\infty\frac{kdk}{2\pi}\left\{ C_s \left[g_s(\omega_n-\Omega_m,\bm k)g_s(\omega_n,\bm k)^3-g_s(\omega_n+\Omega_m,\bm k)g_s(\omega_n,\bm k)^3\right]	\right.\\
&&\left. +{\rm Re}[A_{s}]\left[g_s(\omega_n-\Omega_m, k)g_s(\omega_n, k)g_{\bar{s}}(\omega_n, k)^2 - g_s(\omega_n+\Omega_m, k)g_s(\omega_n, k)g_{\bar{s}}(\omega_n, k)^2\right]\right\}\\
\nonumber &\approx&-\frac{eJ\Omega\tau^2(E_F^2-m_0^2)}{4\pi E_F^2}{\rm sgn}(E_F)\Theta(E_F^2-m_0^2)
\\
\nonumber&&\times\left[\delta_{ik}\delta_{jl}\left(1 -\gamma\frac{E_F^2+3m_0^2}{2E_Fv_F^2}-\lambda^2\frac{(E_F^2-m_0^2)^2(4E_F^6-15m_0^2E_F^4-10m_0^4E_F^2-9m_0^6)}{2E_F^6 v_F^6}\right)\right.\\
\nonumber &&\hspace{0.25cm}+ \delta_{ij}\delta_{kl}\left(3\gamma\frac{(E_F^2-m_0^2)}{2E_Fv_F^2}+9\lambda^2\frac{(E_F^2-m_0^2)(E_F^8+m_0^4E_F^4+m_0^6E_F^2+3m_0^8)}{2E_F^6v_F^6}\right)\\
&&\hspace{0.25cm}\left. -\delta_{il}\delta_{jk}\left(\gamma\frac{(E_F^2-m_0^2)}{2E_Fv_F^2}	+ \lambda^2\frac{3(E_F^2-m_0^2)(2E_F^8-12m_0^2 E_F^6 + 2m_0^4E_F^4+5m_0^6E_F^2-15m_0^8)}{4E_F^6v_F^6}\right)\right]
\eea
where $C_s$ are the matrix elements defined as
\bea
C_s &\equiv& \int_0^{2\pi}\frac{d\theta}{2\pi} \braket{s|v_k|s}\braket{s|\sigma_j|s}\braket{s|v_l|s}\braket{s|\sigma_i|s}.
\eea

Consequently, the electrical spin susceptibility is obtained as
\bea
\chi_i^f &=& -\frac{eJ^2\tau}{2\pi E_F^2} {\rm sgn}(E_F)\Theta(E_F^2-m_0^2) \Lambda_i^f(E_F)\\
\chi_i^d &=& \frac{eJ\tau^2}{4\pi E_F^2} {\rm sgn}(E_F)\Theta(E_F^2-m_0^2) \Lambda_i^d(E_F)
\eea
where
\bea
\Lambda_1^f(E_F) &=& 1- \gamma \frac{2m_0^2}{E_F v_F^2} - \lambda^2\frac{(E_F^2-m_0^2)^3(31E_F^2+9m_0^2)}{8E_F^4 v_F^6},\\
\Lambda_2^f(E_F) &=& \gamma \frac{E_F^2-m_0^2}{E_Fv_F^2}+ \lambda^2\frac{(E_F^2-m_0^2)^2(E_F^6-34m_0^2E_F^4-35m_0^4E_F^2-40m_0^6)}{8E_F^6 v_F^6},\\
\Lambda_3^f(E_F) &=& -\gamma \frac{E_F^2-m_0^2}{E_Fv_F^2} - \lambda^2\frac{(E_F^2-m_0^2)^2(31E_F^6-10m_0^2E_F^4-17m_0^4E_F^2-40m_0^6)}{8E_F^6 v_F^6},\\
\Lambda_1^d(E_F) &=& E_F^2-m_0^2 -\gamma\frac{\left(E_F^2+3m_0^2\right)\left(E_F^2-m_0^2\right)}{2E_Fv_F^2}-\lambda^2\frac{(E_F^2-m_0^2)^3(4E_F^6-15m_0^2E_F^4-10m_0^4E_F^2-9m_0^6)}{2E_F^6 v_F^6},\\
\Lambda_2^d(E_F) &=& 3\gamma\frac{(E_F^2-m_0^2)^2}{2E_Fv_F^2}+9\lambda^2\frac{(E_F^2-m_0^2)^2(E_F^8+m_0^4E_F^4+m_0^6E_F^2+3m_0^8)}{2E_F^6v_F^6},\\
\Lambda_3^d(E_F) &=& -\gamma\frac{(E_F^2-m_0^2)^2}{2E_Fv_F^2}	 - \lambda^2\frac{3(E_F^2-m_0^2)^2(2E_F^8-12m_0^2 E_F^6 + 2m_0^4E_F^4+5m_0^6E_F^2-15m_0^8)}{4E_F^6v_F^6}.
\eea
In the limit of $|E_F|>>|JM_z|$, the Eq.(14) and (15) are obtained.
\end{widetext}

%

\end{document}